\documentclass[cameraready]{Interspeech}
\title{Transcript-Free Flow-Matching Text-to-Speech \\ via Speech Feature Conditioning}

\author[affiliation={1}]{SooHwan}{Eom}
\author[affiliation={1}]{Hee Suk}{Yoon}
\author[affiliation={1}]{Eunseop}{Yoon}
\author[affiliation={2}]{Mark}{Hasegawa-Johnson}
\author[affiliation={1}, correspondingauthor]{Chang D.}{Yoo}

\address{
    $^1$ Korea Advanced Institute of Science and Technology, South Korea \\
    $^2$ University of Illinois Urbana-Champaign, United States
}

\email{\{sean1105, hskyoon, esyoon97, cd\_yoo\}@kaist.ac.kr, jhasegaw@illinois.edu}

\keywords{text-to-speech, flow-matching, self-supervised learning, dysarthric speech, zero-shot voice cloning}

\usepackage{comment}
\usepackage{amsbsy}
\usepackage{amsmath}
\usepackage{amssymb}
\usepackage{mathtools}
\usepackage{multirow}

\begin{document}

\maketitle

\begin{abstract}
Recent flow-matching text-to-speech (TTS) models, such as F5-TTS, rely on a reference transcript at inference time, obtained from an external ASR system. This dependency makes zero-shot TTS brittle for accented or dysarthric speakers, precisely the scenarios where it is most needed. Moreover, we find that text-based reference conditioning can propagate atypical acoustic patterns from atypical speech into synthesis, even when ground-truth transcripts are available. To address this, we propose RTFree-F5, which replaces the reference transcript with continuous self-supervised speech representations mapped into F5-TTS's text-conditioning space via a lightweight adapter, while reusing the pretrained checkpoint. On dysarthric speech, RTFree-F5 reduces WER from 24.6\% to 10.4\%, surpassing even the ground-truth reference transcript baselines, while improving naturalness and remaining competitive on standard benchmarks without requiring any reference transcript.
\end{abstract}

\section{Introduction}

Zero-shot text-to-speech (TTS) aims to synthesize natural speech for speakers unseen during training, imitating an arbitrary speaker's voice from a short reference sample without further training \cite{valle, valle2, voicebox, ns2, e2tts, f5tts}. A particularly compelling application is \emph{atypical speech reconstruction}: synthesizing intelligible, natural-sounding speech that preserves the identity of speakers with dysarthria, strong accents, or other non-standard speech patterns \cite{diffdsr, unitdsr, accent}. For these speakers, zero-shot TTS could substantially improve communication; yet, they are precisely the cases where current systems are most fragile.

Autoregressive (AR) TTS models, such as neural codec language models, achieve strong zero-shot performance by predicting discrete speech tokens sequentially, enabling implicit duration modeling and flexible sampling \cite{valle, valle2}. However, AR models suffer from exposure bias, high inference latency, and a heavy dependence on the quality of the speech tokenizer.

Non-autoregressive (NAR) TTS models have recently emerged as competitive alternatives, offering fast, parallel generation. With diffusion \cite{ddpm, score} and flow-matching \cite{cfm-ot} approaches, NAR-based systems can now achieve high-fidelity, zero-shot synthesis \cite{voicebox, ns2, voiceflow, matchatts, dittotts}. A major challenge for these models is text-speech alignment: parallel generation requires inferring phoneme durations to align text with the much longer mel-spectrogram. Consequently, many NAR systems rely on a grapheme-to-phoneme (G2P) front-end, a duration predictor, and a phoneme-aligned corpus for supervision.

E2-TTS \cite{e2tts} and F5-TTS \cite{f5tts} sidestep these components by operating directly on character sequences padded with filler tokens to match the mel length. E2-TTS demonstrated that simple character-level conditioning with filler tokens can already achieve human-level zero-shot TTS. F5-TTS further improves text-speech alignment by refining the padded character sequence with an additional text encoder.

However, both models rely on in-context infilling and thus require a transcript of the reference audio at inference time, typically obtained via an external ASR system. For speech with poor articulation, strong accents, or dysarthria, ASR errors directly degrade the reference conditioning. More fundamentally, we show that even with perfect oracle transcripts, text-based reference conditioning can struggle on atypical speech because the infilling mechanism may propagate atypical acoustic patterns from the reference mel-spectrogram into the generated output. Recent work \cite{ezvc} proposed a textless, zero-shot voice conversion framework that reuses the F5-TTS architecture and conditions it on discrete self-supervised speech units rather than text. However, because the input space is no longer characters but unit IDs, the original character-based F5-TTS checkpoint cannot be reused and must be trained from scratch.

\begin{figure*}[t]
    \centering
    \includegraphics[width=\linewidth]{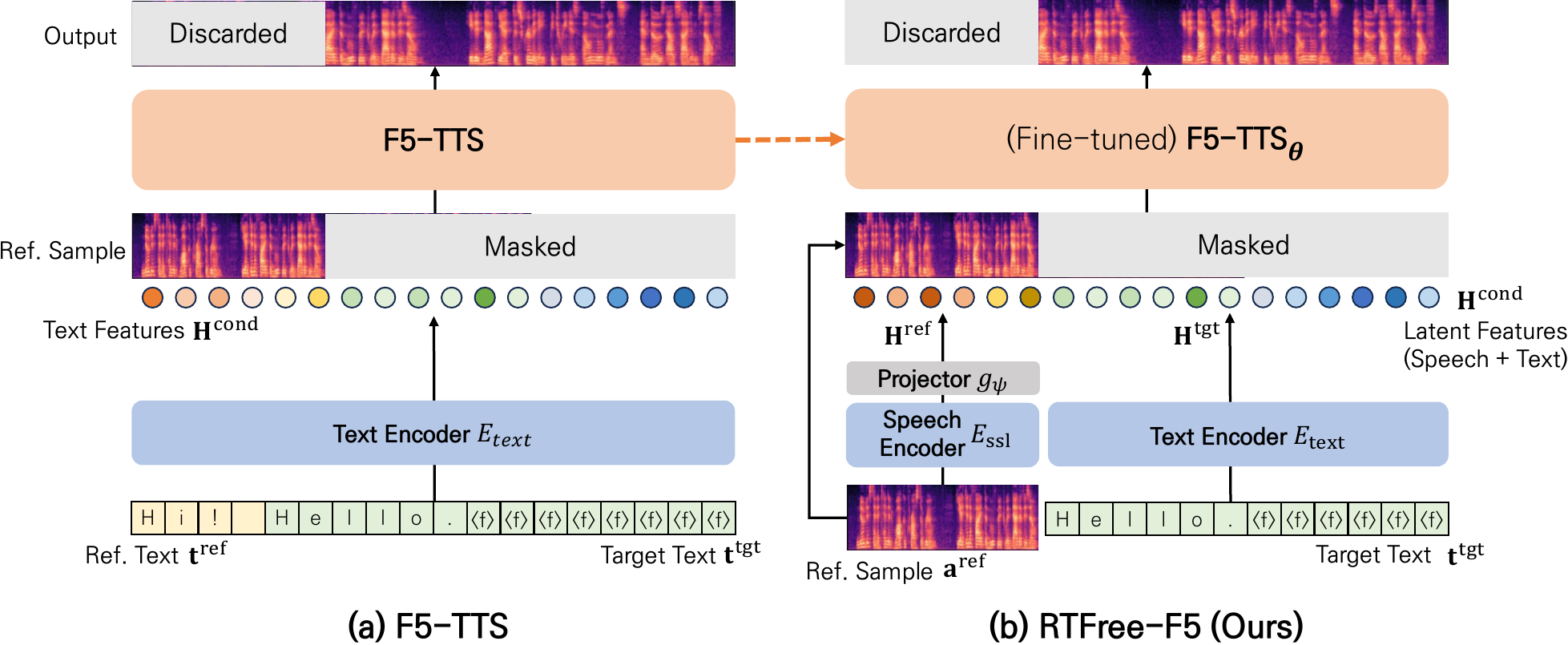}
    \caption{Comparison between F5-TTS (left) and RTFree-F5 (right). F5-TTS conditions on the concatenated reference and target transcripts $[\mathbf{t}^{\mathrm{ref}}; \mathbf{t}^{\mathrm{tgt}}]$ via a shared text encoder $E_{\mathrm{text}}$. RTFree-F5 replaces the reference transcript with self-supervised speech features extracted by $E_{\mathrm{ssl}}$ and projected into the text-conditioning space via $g_\psi$, eliminating the need for a reference transcript at inference.}
    \label{fig:overview}
\end{figure*}

To address these limitations, we propose \textbf{RTFree-F5}, a reference-guided TTS framework that (1) removes the dependence on reference transcripts by conditioning on continuous self-supervised speech features, while (2) fully reusing the pretrained F5-TTS checkpoint. Rather than discretizing speech features into new tokens, RTFree-F5 maps continuous speech representations into the existing F5-TTS conditioning space via a lightweight projector. The frozen speech encoder and projector together produce ``latent text'' features from the reference audio without its transcript, while the original text encoder still conditions on the target transcript. This speech-text condition alignment removes the need for reference transcripts at inference time, preserves text-based content control, and fully reuses the pretrained checkpoint.

We train RTFree-F5 on cross-utterance pairs from the same speaker in a two-stage strategy: first aligning the projector to the frozen F5-TTS conditioning space, then jointly fine-tuning the projector and flow-matching backbone. We validate our framework on standard zero-shot TTS benchmarks as well as more challenging non-native TTS and identity-preserving dysarthric speech reconstruction scenarios, demonstrating that SSL-based conditioning substantially outperforms text-based conditioning for atypical speakers.

\section{Method}

We propose \textbf{RTFree-F5} (\textbf{R}eference \textbf{T}ranscript-\textbf{Free} \textbf{F5}-TTS), which extends F5-TTS \cite{f5tts} by replacing its text-based reference conditioning with self-supervised speech representations, eliminating the need for a reference transcript at inference time. Figure~\ref{fig:overview} illustrates the overall architecture. We first review the F5-TTS conditioning mechanism, then describe our approach.

\subsection{Background: F5-TTS Conditioning}

F5-TTS \cite{f5tts} is a non-autoregressive TTS model that uses flow-matching with a Diffusion Transformer (DiT) \cite{dit} backbone. It eliminates conventional TTS components (G2P, duration prediction, forced alignment) by tokenizing input text at the character level and padding with filler tokens to match the target mel-spectrogram length. A ConvNeXt V2 \cite{convnextv2} text encoder $E_{\mathrm{text}}$ processes these tokens into conditioning features. The model is trained with an infilling objective: a contiguous segment of the mel-spectrogram is masked, and the DiT learns to reconstruct it conditioned on the unmasked speech context and the full transcription. At inference time, the unmasked portion serves as the acoustic prompt for zero-shot voice cloning.

In this framework, both the reference transcript and the target text are tokenized, padded, and jointly encoded:
\begin{equation}
    \mathbf{H}^{\mathrm{cond}} = E_{\mathrm{text}}\left( \left[ \mathbf{t}^{\mathrm{ref}};\, \mathbf{t}^{\mathrm{tgt}} \right] \right) \in \mathbb{R}^{(T_{\mathrm{ref}} + T_{\mathrm{tgt}}) \times D},
\end{equation}
where $\mathbf{t}^{\mathrm{ref}}$ and $\mathbf{t}^{\mathrm{tgt}}$ denote padded token sequences for the reference and target transcripts, and $D$ is the embedding dimension. This design requires the reference transcript at both training and inference time.

\subsection{Self-Supervised Speech Representations}
To remove the dependency on reference transcripts, we replace text-based reference features with self-supervised learning (SSL) speech representations. Given a reference waveform $\mathbf{a}^{\mathrm{ref}}$, we extract frame-level representations using a frozen pre-trained SSL encoder $E_{\mathrm{ssl}}$:
\begin{equation}
    \mathbf{H}^{\mathrm{ssl}} = E_{\mathrm{ssl}}(\mathbf{a}^{\mathrm{ref}}) \in \mathbb{R}^{T_{\mathrm{ref}} \times D_{\mathrm{ssl}}},
\end{equation}
where $T_{\mathrm{ref}}$ is the number of frames and $D_{\mathrm{ssl}}$ is the SSL embedding dimension. We adopt WavLM \cite{wavlm} as our speech encoder, as it captures both content and speaker characteristics in its intermediate representations.

\subsection{Modality Bridging via Projection}
Since the SSL representations reside in a different embedding space than the F5-TTS text features, we introduce a projection module $g_\psi$ to bridge these two modalities. We employ a multi-layer perceptron (MLP) that maps SSL features to the text-conditioning space:
\begin{equation}
    \mathbf{H}^{\mathrm{ref}} = g_\psi(\mathbf{H}^{\mathrm{ssl}}) \in \mathbb{R}^{T_{\mathrm{ref}} \times D},
\end{equation}
where $g_\psi: \mathbb{R}^{D_{\mathrm{ssl}}} \rightarrow \mathbb{R}^{D}$ is applied independently to each frame. The projector is a two-layer MLP with LayerNorm \cite{ln}:
\begin{equation}
g_\psi(\mathbf{h}) = \text{LayerNorm}\left(\mathbf{W}_2 \cdot \sigma(\mathbf{W}_1 \mathbf{h} + \mathbf{b}_1) + \mathbf{b}_2\right),
\end{equation}
where $\sigma(\cdot)$ denotes a non-linear activation function (e.g., GELU \cite{gelu}), and $\psi = \{\mathbf{W}_1, \mathbf{b}_1, \mathbf{W}_2, \mathbf{b}_2\}$ are learnable parameters.

\subsection{Conditioning Mechanism}
The projected reference features $\mathbf{H}^{\mathrm{ref}}$ replace the text-encoded reference representations in the original F5-TTS conditioning. The target text is still processed through the text encoder:
\begin{equation}
    \mathbf{H}^{\mathrm{tgt}} = E_{\mathrm{text}}(\mathbf{t}^{\mathrm{tgt}}) \in \mathbb{R}^{T_{\mathrm{tgt}} \times D}.
\end{equation}

The final conditioning input is formed by concatenating the projected speech features with the target text features along the temporal axis:
\begin{equation}
    \mathbf{H}^{\mathrm{cond}} = \left[ \mathbf{H}^{\mathrm{ref}};  \mathbf{H}^{\mathrm{tgt}} \right] \in \mathbb{R}^{(T_{\mathrm{ref}} + T_{\mathrm{tgt}}) \times D}.
\end{equation}

This formulation maintains compatibility with the original F5-TTS architecture, where the DiT backbone receives the concatenated conditioning alongside the masked mel-spectrogram input. Crucially, only the \emph{conditioning channel} changes from the original F5-TTS: the reference mel-spectrogram remains the unmasked acoustic context, as in the original pipeline.

\subsection{Training}
\label{sec:training}
We adopt a two-stage training strategy, progressively adapting the model from cross-modal alignment to joint generative fine-tuning. Unlike the original F5-TTS infilling setup, which partitions a single utterance into (unmasked) reference and (masked) target segments, we train on \textbf{cross-utterance pairs}: the reference and target are distinct utterances from the same speaker. This resembles the inference scenario and encourages the model to learn task-specific, robust speaker representations rather than relying on local acoustic continuity.

\subsubsection{Stage 1: Cross-Modal Alignment}
In the first stage, we train only the projection module $g_\psi$ while keeping both the SSL encoder $E_{\mathrm{ssl}}$ and the entire F5-TTS model frozen. The goal is to learn an initial alignment between SSL speech representations and the F5-TTS text-conditioning space.

Let $(\mathbf{a}^{\mathrm{ref}}, \mathbf{x}^{\mathrm{tgt}}, \mathbf{t}^{\mathrm{tgt}})$ denote a training tuple, where $\mathbf{a}^{\mathrm{ref}}$ is the reference audio, $\mathbf{x}^{\mathrm{tgt}}$ is the target mel-spectrogram, and $\mathbf{t}^{\mathrm{tgt}}$ is the target transcription. The reference and target are sampled from different utterances of the same speaker. The conditioning for F5-TTS is constructed as:
\begin{equation}
    \mathbf{H}^{\mathrm{cond}}_\psi = \left[ g_\psi\left( E_{\mathrm{ssl}}(\mathbf{a}^{\mathrm{ref}}) \right);\, E_{\mathrm{text}}(\mathbf{t}^{\mathrm{tgt}}) \right].
\end{equation}

The projector parameter $\psi$ is optimized via the flow-matching objective:
\begin{equation}
    \mathcal{L}_{\mathrm{stage1}}(\psi) = \mathbb{E}_{t,\, \mathbf{x}^{\mathrm{tgt}},\, \boldsymbol{\epsilon}} \left\Vert \mathbf{v} - \mathbf{v}_\theta\left(\mathbf{z}_t, t, \mathbf{H}^{\mathrm{cond}}_\psi\right) \right\Vert^2,
\end{equation}
where $\mathbf{z}_t = (1-t)\boldsymbol{\epsilon} + t\mathbf{x}^{\mathrm{tgt}}$, $\mathbf{v} = \mathbf{x}^{\mathrm{tgt}} - \boldsymbol{\epsilon}$, $\boldsymbol{\epsilon}\sim\mathcal{N}(0,\mathbf{I})$. Here, $v_\theta$ denotes the DiT backbone parameterized by $\theta$. During Stage~1, $\theta$ is kept frozen and gradients are backpropagated only through the projector.

\subsubsection{Stage 2: Joint Fine-tuning}
In the second stage, we jointly fine-tune the projector $g_\psi$ and the DiT backbone $\mathbf{v}_\theta$, while keeping both the text and speech encoders frozen. This stage is necessary because optimizing only the projector in Stage 1 may be insufficient, potentially leading to a distribution mismatch for the DiT backbone. Moreover, the pretrained DiT was originally trained under a within-utterance infilling objective, whereas our task involves cross-utterance conditioning, in which the reference and target speech are acoustically disjoint. Joint fine-tuning enables the model to adapt to this new conditioning regime.

The flow-matching training objective remains the same:
\begin{equation}
    \mathcal{L}_{\mathrm{stage2}}(\psi, \theta) = \mathbb{E}_{t,\, \mathbf{x}^{\mathrm{tgt}},\, \boldsymbol{\epsilon}} \left\Vert \mathbf{v} - \mathbf{v}_\theta\left(\mathbf{z}_t, t, \mathbf{H}^{\mathrm{cond}}_\psi\right) \right\Vert^2,
\end{equation}
with gradients now flowing through both the projector and DiT.

\subsection{Inference}
At inference, given reference audio $\mathbf{a}^{\mathrm{ref}}$ and target text $\mathbf{t}^{\mathrm{tgt}}$, we construct $\mathbf{H}^{\mathrm{cond}} = [g_\psi(E_{\mathrm{ssl}}(\mathbf{a}^{\mathrm{ref}}));\, E_{\mathrm{text}}(\mathbf{t}^{\mathrm{tgt}})]$ and solve the ODE $d\mathbf{z}_t / dt = \mathbf{v}_\theta(\mathbf{z}_t, t, \mathbf{H}^{\mathrm{cond}})$ from $\mathbf{z}_0 \sim \mathcal{N}(\mathbf{0}, \mathbf{I})$ to $t = 1$ using classifier-free guidance (CFG)~\cite{cfg}. The resulting mel-spectrogram $\mathbf{z}_1$ is passed through a vocoder to synthesize the final waveform. 
No transcription of the reference audio is required during inference time.

\section{Experimental Setup}

\subsection{Implementation Details}

We build RTFree-F5 upon the pretrained F5-TTS v1 Base checkpoint\footnote{\url{https://huggingface.co/SWivid/F5-TTS}}. The SSL speech encoder is WavLM-Large\footnote{\url{https://huggingface.co/microsoft/wavlm-large}}, which remains frozen throughout training. The cross-modal projector is a two-layer MLP that maps 1024-dimensional WavLM features to the 512-dimensional F5-TTS conditioning space, with the hidden dimension also set to 512, resulting in 0.8M trainable parameters. For training stability, we initialize the final projection layer with a weight scaling factor of 0.1.

Since WavLM operates at 50\,Hz while F5-TTS mel-spectrograms use 24\,kHz audio with 256-sample hop length ($\approx$93.75\,Hz), we apply linear interpolation to upsample the projected speech features to the mel-spectrogram frame rate.

\subsection{Training}

We use the two-stage training pipeline described in Section~\ref{sec:training}. Stage~1 is trained for 10 epochs, followed by Stage~2 for 20 epochs. We use the AdamW optimizer with linear warmup followed by linear decay. To preserve the pretrained linguistic prior of the DiT backbone and avoid overfitting, we use a lower learning rate of $1\times10^{-5}$ for the backbone and a higher rate of $5\times10^{-5}$ for the projector. All experiments are conducted on 4 NVIDIA A100 GPUs; Stage~1 requires approximately 1-2 days of training, while Stage~2 takes about 2-3 days. 

For training data construction, we sample cross-utterance pairs from the same speaker in the LibriTTS \cite{libritts} training set. We filter samples to have durations between 0.3 and 30 seconds. Following F5-TTS, we apply CFG dropout with an audio-drop probability of 0.3 and a joint-drop probability of 0.2 \cite{f5tts}.

\subsection{Inference}

Following the default settings from F5-TTS, we use the Euler ODE solver with 32 function evaluations (NFE), a sway sampling coefficient of $-1$, and a CFG strength of $2.0$ \cite{f5tts}. We use the Vocos \cite{vocos} vocoder to convert generated mel-spectrograms to waveforms at 24\,kHz.

\subsection{Evaluation Datasets}

We evaluate on both typical- and atypical-speaker datasets. For \emph{typical speakers}, we use \texttt{test-clean} split of \textbf{LibriSpeech-PC} \cite{librispeechpc}  and \texttt{test-en} split of \textbf{SeedTTS} \cite{seedtts} . For \emph{atypical speakers}, we use the \texttt{dev} split of \textbf{SAP (Speech Accessibility Project)} \cite{sap}, which contains dysarthric speech from roughly 100 speakers, and \textbf{L2-ARCTIC} \cite{l2arctic}, which provides accented English from non-native speakers across six L1 backgrounds. We select two speakers per language for evaluation.

\subsection{Metrics and Baselines}

We evaluate three metrics for synthesis quality: word error rate (\textbf{WER}) of Whisper large-v3 \cite{whisper} transcripts for intelligibility, speaker similarity (\textbf{SIM}) via ECAPA-TDNN \cite{ecapa} feature cosine similarity for speaker identity, and predicted naturalness (\textbf{MOS}) via UTMOS \cite{utmos}.
We compare against F5-TTS baselines using \textbf{oracle} transcripts and Whisper large-v3 \textbf{ASR} transcripts. 
For atypical speakers, we additionally report the WER, SIM, and MOS of the \textbf{original} speech as a reference point, with SIM computed against utterances from the same speaker.

\section{Results}

\subsection{Typical Speaker Evaluation}

Table~\ref{tab:typical} presents results on standard zero-shot TTS benchmarks with typical speakers. 

\begin{table}[t]
\centering
\caption{Zero-shot TTS results on \textbf{typical}-speaker benchmarks. \textbf{Baseline} refers to the pretrained F5-TTS model conditioned on either \textbf{oracle} or \textbf{ASR} transcripts. \textbf{WER} denotes word error rate, \textbf{SIM} denotes speaker similarity, and \textbf{MOS} denotes naturalness predicted by UTMOS \cite{utmos}.}
\label{tab:typical}
\resizebox{\columnwidth}{!}{%
\begin{tabular}{l|ccc|ccc}
\toprule
\multirow{2}{*}{Model} & \multicolumn{3}{c|}{LibriSpeech-PC} & \multicolumn{3}{c}{SeedTTS} \\
& WER$\downarrow$ & SIM$\uparrow$ & MOS$\uparrow$ & WER$\downarrow$ & SIM$\uparrow$ & MOS$\uparrow$ \\
\midrule
Baseline (oracle) & 2.08\% & 0.67 & 3.83 & \textbf{1.43\%} & \textbf{0.68} & 3.66 \\
Baseline (ASR) & 2.17\% & \textbf{0.68} & 3.84 & 1.45\% & \textbf{0.68} & 3.66 \\
\midrule
RTFree Stage 1 & 4.68\% & 0.64 & 3.91 & 2.86\% & 0.62 & 3.80 \\
\textbf{RTFree Stage 2} & \textbf{1.77\%} & 0.66 & \textbf{4.13} & 1.56\% & 0.63 & \textbf{3.94} \\
\bottomrule
\end{tabular}%
}
\end{table}

On LibriSpeech-PC, RTFree-F5 (Stage~2) achieves a WER of 1.77\%, outperforming both the oracle baseline (2.08\%) and ASR baseline (2.17\%). The MOS improves substantially from 3.83 to 4.13, indicating improved naturalness. Speaker similarity remains comparable (0.66 vs.\ 0.67). Similar trends are observed on SeedTTS, where RTFree-F5 achieves the best MOS (3.94) with competitive WER (1.56\%).

The Stage~1 model, which trained only the projector, shows degraded performance (4.68\% WER on LibriSpeech-PC), confirming that joint fine-tuning in Stage~2 is essential for the model to effectively utilize speech-based conditioning.

\subsection{Atypical Speaker Evaluation}

Table~\ref{tab:atypical} presents results on dysarthric (SAP) and non-native (L2-ARCTIC) speakers, the primary motivation for our method.

\begin{table}[t]
\centering
\caption{Zero-shot TTS results on \textbf{atypical}-speaker benchmarks. \textbf{Original} reports reference metrics computed on the unmodified source recordings.}
\label{tab:atypical}
\resizebox{\columnwidth}{!}{%
\begin{tabular}{l|ccc|ccc}
\toprule
\multirow{2}{*}{Model} & \multicolumn{3}{c|}{SAP (Dysarthric)} & \multicolumn{3}{c}{L2-ARCTIC (Non-native)} \\
& WER$\downarrow$ & SIM$\uparrow$ & MOS$\uparrow$ & WER$\downarrow$ & SIM$\uparrow$ & MOS$\uparrow$ \\
\midrule
Original & 24.62\% & 0.71 & 2.16 & 10.75\% & 0.73 & 3.82 \\
\midrule
Baseline (oracle) & 20.71\% & \textbf{0.60} & 2.27 & 2.00\% & 0.59 & 3.92 \\
Baseline (ASR) & 20.46\% & \textbf{0.60} & 2.27 & 1.99\% & 0.60 & 3.92 \\
\midrule
RTFree Stage 1 & 90.00\% & 0.52 & 2.19 & 7.53\% & 0.49 & 4.00 \\
\textbf{RTFree Stage 2} & \textbf{10.39\%} & 0.50 & \textbf{2.85} & \textbf{1.44\%} & \textbf{0.61} & \textbf{4.08} \\
\bottomrule
\end{tabular}%
}
\end{table}

\noindent\textbf{Dysarthric Speakers (SAP):} RTFree-F5 achieves notable improvements in intelligibility. The WER drops from 24.62\% (original dysarthric speech) to 10.39\%, a \textbf{58\% relative reduction}. Remarkably, RTFree-F5 also outperforms the oracle baseline (20.71\% WER), which has access to perfect reference transcripts. Moreover, the MOS improves from 2.16 (original) to 2.85, representing a notable gain in perceived naturalness.

At the same time, speaker similarity decreases from 0.60 to 0.50, suggesting a trade-off between intelligibility and identity preservation. This may occur because improving intelligibility requires partially correcting atypical articulatory patterns that are intrinsically linked to the speaker's identity. Moreover, ECAPA-TDNN embeddings may conflate speaker identity with pathological speech characteristics; dysarthric speech that is partially normalized may remain perceptually identifiable despite a lower embedding similarity score. In accessibility applications, where effective communication is the primary goal, such a trade-off may be acceptable, though improving speaker preservation remains an important direction for future work.

The Stage~1 model catastrophically fails on SAP, yielding a WER of 90\%, which reveals the severity of the distribution shift between dysarthric and healthy speech. This underscores the necessity of Stage~2 fine-tuning to improve the generalizability of DiT backbone under atypical speech conditioning.

\noindent\textbf{Non-native Speakers (L2-ARCTIC):} RTFree-F5 reduces WER from 10.75\% (original) to 1.44\%, an \textbf{87\% relative reduction}. Notably, RTFree-F5 also outperforms both the oracle baseline (2.00\%) and the ASR baseline (1.99\%), demonstrating that SSL-based conditioning can surpass text-based conditioning even when accurate transcripts are available. Speaker similarity is maintained (0.61 vs.\ 0.60), while naturalness (MOS) improves from 3.82 (original) to 4.08.

\subsection{Analysis}
One key finding is that the oracle baseline, despite access to ground-truth transcripts, underperforms on both atypical speaker datasets. This suggests a crucial limitation of text-based reference conditioning that extends beyond ASR reliability.

Both the oracle baseline and RTFree-F5 use the reference mel-spectrogram as the unmasked acoustic context in the infilling setup; at inference time, the reference-conditioning path differs in the conditioning channel. In the oracle baseline, the conditioning concatenates text features from the reference transcript with those from the target text. These text features encode an expected phonetic sequence that may be inconsistent with the actual acoustic patterns in the reference mel-spectrogram. For dysarthric speakers, this mismatch appears as irregular timing, unclear pronunciation, or atypical prosody; for non-native speakers, it can appear as accented pronunciations and speech rhythms that differ from native patterns. In both cases, the DiT must handle conflicting information between the text conditionings and acoustic contexts, and the resulting synthesis tends to propagate atypical patterns from the reference mel-spectrogram into the generated speech.

RTFree-F5 instead replaces reference text features with projected SSL representations extracted from the same reference audio, producing a conditioning signal that is intrinsically aligned with the underlying acoustics. Text features encode what sounds should be produced, which may differ from what the speaker actually produces. On the other hand, SSL representations capture how the speaker sounds without imposing such expectations. As a result, the DiT receives conditioning that is compatible with the reference mel-spectrogram while being less tied to specific pronunciation patterns. Consequently, the flow-matching decoder can preserve speaker identity from the reference while generating speech with more typical pronunciation, guided primarily by the target text.

\section{Conclusion}

We presented RTFree-F5, a framework that eliminates reference transcript dependency in flow-matching TTS by projecting continuous WavLM features into the text-conditioning space of a pretrained F5-TTS model via a lightweight MLP projector. Our experiments reveal that text-based reference conditioning struggles with atypical speech, due to a mismatch between normative text features and pathological acoustic content. RTFree-F5 reduces dysarthric speech WER from 24.6\% to 10.4\% with substantial naturalness gains, demonstrating the efficacy of SSL-based conditioning for atypical speakers.

\section{Generative AI Use Disclosure}

Large Language Models were used exclusively to correct grammar and refine the wording of the manuscript text. No original ideas, analyses, or passages were generated by these tools. All authors reviewed AI-assisted edits and accept full responsibility for the final manuscript.

\section{Acknowledgements}
This work was supported by Institute of Information \& communications Technology Planning \& Evaluation (IITP) grant funded by the Korea government(MSIT) (No.RS-2022-II220184, Development and Study of AI Technologies to Inexpensively Conform to Evolving Policy on Ethics) and Institute for Information \& communications Technology Planning \& Evaluation (IITP) grant funded by the Korea government(MSIT) (No.RS-2021-II211381, Development of Causal AI through Video Understanding and Reinforcement Learning, and Its Applications to Real Environments).

\bibliographystyle{IEEEtran}
\bibliography{main}

@inproceedings{f5tts,
  title={F5-tts: A fairytaler that fakes fluent and faithful speech with flow matching},
  author={Chen, Yushen and Niu, Zhikang and Ma, Ziyang and Deng, Keqi and Wang, Chunhui and JianZhao, JianZhao and Yu, Kai and Chen, Xie},
  booktitle={Proceedings of the 63rd Annual Meeting of the Association for Computational Linguistics (Volume 1: Long Papers)},
  pages={6255--6271},
  year={2025}
}

@inproceedings{librispeechpc,
  title={{LibriSpeech-PC}: Benchmark for Evaluation of Punctuation and Capitalization Capabilities of end-to-end {ASR} Models},
  author={Meister, Aleksandr and Novikov, Matvei and Karpov, Nikolay and Bakhturina, Evelina and Lavrukhin, Vitaly and Ginsburg, Boris},
  booktitle={Proc. ASRU},
  pages={1--7},
  year={2023},
  organization={IEEE}
}

@inproceedings{mos,
  title={Crowdmos: An approach for crowdsourcing mean opinion score studies},
  author={Ribeiro, Fl{\'a}vio and Flor{\^e}ncio, Dinei and Zhang, Cha and Seltzer, Michael},
  booktitle={2011 IEEE international conference on acoustics, speech and signal processing (ICASSP)},
  pages={2416--2419},
  year={2011},
  organization={IEEE}
}

@inproceedings{whisper,
  title={Robust speech recognition via large-scale weak supervision},
  author={Radford, Alec and Kim, Jong Wook and Xu, Tao and Brockman, Greg and McLeavey, Christine and Sutskever, Ilya},
  booktitle={International conference on machine learning},
  pages={28492--28518},
  year={2023},
  organization={PMLR}
}

@article{wavlm,
   title={WavLM: Large-Scale Self-Supervised Pre-Training for Full Stack Speech Processing},
   volume={16},
   ISSN={1941-0484},
   url={http://dx.doi.org/10.1109/JSTSP.2022.3188113},
   DOI={10.1109/jstsp.2022.3188113},
   number={6},
   journal={IEEE Journal of Selected Topics in Signal Processing},
   publisher={Institute of Electrical and Electronics Engineers (IEEE)},
   author={Chen, Sanyuan and Wang, Chengyi and Chen, Zhengyang and Wu, Yu and Liu, Shujie and Chen, Zhuo and Li, Jinyu and Kanda, Naoyuki and Yoshioka, Takuya and Xiao, Xiong and Wu, Jian and Zhou, Long and Ren, Shuo and Qian, Yanmin and Qian, Yao and Wu, Jian and Zeng, Michael and Yu, Xiangzhan and Wei, Furu},
   year={2022},
   month=oct, pages={1505–1518} }

@inproceedings{
vocos,
title={Vocos: Closing the gap between time-domain and Fourier-based neural vocoders for high-quality audio synthesis},
author={Hubert Siuzdak},
booktitle={The Twelfth International Conference on Learning Representations},
year={2024},
url={https://openreview.net/forum?id=vY9nzQmQBw}
}

@inproceedings{convnextv2,
  title={Convnext v2: Co-designing and scaling convnets with masked autoencoders},
  author={Woo, Sanghyun and Debnath, Shoubhik and Hu, Ronghang and Chen, Xinlei and Liu, Zhuang and Kweon, In So and Xie, Saining},
  booktitle={Proceedings of the IEEE/CVF Conference on Computer Vision and Pattern Recognition},
  pages={16133--16142},
  year={2023}
}

@inproceedings{dit,
  title={Scalable diffusion models with transformers},
  author={Peebles, William and Xie, Saining},
  booktitle={Proceedings of the IEEE/CVF International Conference on Computer Vision},
  pages={4195--4205},
  year={2023}
}

@article{cfg,
  title={Classifier-free diffusion guidance},
  author={Ho, Jonathan and Salimans, Tim},
  journal={arXiv preprint arXiv:2207.12598},
  year={2022}
}

@article{seedtts,
  title={{Seed-TTS}: A Family of High-Quality Versatile Speech Generation Models},
  author={Anastassiou, Philip and Chen, Jiawei and Chen, Jitong and Chen, Yuanzhe and others},
  journal={arXiv preprint arXiv:2406.02430},
  year={2024}
}

@inproceedings{e2tts,
  title={E2 tts: Embarrassingly easy fully non-autoregressive zero-shot tts},
  author={Eskimez, Sefik Emre and Wang, Xiaofei and Thakker, Manthan and Li, Canrun and Tsai, Chung-Hsien and Xiao, Zhen and Yang, Hemin and Zhu, Zirun and Tang, Min and Tan, Xu and others},
  booktitle={2024 IEEE spoken language technology workshop (SLT)},
  pages={682--689},
  year={2024},
  organization={IEEE}
}

@inproceedings{
dittotts,
title={Di{TT}o-{TTS}: Diffusion Transformers for Scalable Text-to-Speech without Domain-Specific Factors},
author={Keon Lee and Dong Won Kim and Jaehyeon Kim and Seungjun Chung and Jaewoong Cho},
booktitle={The Thirteenth International Conference on Learning Representations},
year={2025},
url={https://openreview.net/forum?id=hQvX9MBowC}
}

@inproceedings{matchatts,
  title={{Matcha-TTS}: A fast {TTS} architecture with conditional flow matching},
  author={Mehta, Shivam and Tu, Ruibo and Beskow, Jonas and Sz{\'e}kely, {\'E}va and Henter, Gustav Eje},
  booktitle={Proc. ICASSP},
  pages={11341--11345},
  year={2024},
  organization={IEEE}
}

@inproceedings{voiceflow,
  title={Voiceflow: Efficient text-to-speech with rectified flow matching},
  author={Guo, Yiwei and Du, Chenpeng and Ma, Ziyang and Chen, Xie and Yu, Kai},
  booktitle={Proc. ICASSP},
  pages={11121--11125},
  year={2024},
  organization={IEEE}
}

@inproceedings{
cfm-ot,
title={Flow Matching for Generative Modeling},
author={Yaron Lipman and Ricky T. Q. Chen and Heli Ben-Hamu and Maximilian Nickel and Matthew Le},
booktitle={The Eleventh International Conference on Learning Representations },
year={2023},
url={https://openreview.net/forum?id=PqvMRDCJT9t}
}

@inproceedings{score,
  author       = {Yang Song and
                  Jascha Sohl{-}Dickstein and
                  Diederik P. Kingma and
                  Abhishek Kumar and
                  Stefano Ermon and
                  Ben Poole},
  title        = {Score-Based Generative Modeling through Stochastic Differential Equations},
  booktitle    = {9th International Conference on Learning Representations, {ICLR} 2021,
                  Virtual Event, Austria, May 3-7, 2021},
  publisher    = {OpenReview.net},
  year         = {2021},
  url          = {https://openreview.net/forum?id=PxTIG12RRHS},
  bibsource    = {dblp computer science bibliography, https://dblp.org}
}

@article{ddpm,
  title={Denoising diffusion probabilistic models},
  author={Ho, Jonathan and Jain, Ajay and Abbeel, Pieter},
  journal={Advances in neural information processing systems},
  volume={33},
  pages={6840--6851},
  year={2020}
}

@article{voicebox,
  title={Voicebox: Text-guided multilingual universal speech generation at scale},
  author={Le, Matthew and Vyas, Apoorv and Shi, Bowen and Karrer, Brian and others},
  journal={Advances in neural information processing systems},
  volume={36},
  year={2024}
}

@article{valle2,
  title={{VALL-E 2}: Neural Codec Language Models are Human Parity Zero-Shot Text to Speech Synthesizers},
  author={Chen, Sanyuan and Liu, Shujie and Zhou, Long and Liu, Yanqing and Tan, Xu and Li, Jinyu and Zhao, Sheng and Qian, Yao and Wei, Furu},
  journal={arXiv preprint arXiv:2406.05370},
  year={2024}
}

@inproceedings{
ns2,
title={NaturalSpeech 2: Latent Diffusion Models are Natural and Zero-Shot Speech and Singing Synthesizers},
author={Kai Shen and Zeqian Ju and Xu Tan and Eric Liu and Yichong Leng and Lei He and Tao Qin and sheng zhao and Jiang Bian},
booktitle={The Twelfth International Conference on Learning Representations},
year={2024},
url={https://openreview.net/forum?id=Rc7dAwVL3v}
}

@article{valle,
  title={Neural codec language models are zero-shot text to speech synthesizers},
  author={Chen, Sanyuan and Wang, Chengyi and Wu, Yu and Zhang, Ziqiang and Zhou, Long and Liu, Shujie and Chen, Zhuo and Liu, Yanqing and Wang, Huaming and Li, Jinyu and others},
  journal={IEEE Transactions on Audio, Speech and Language Processing},
  volume={33},
  pages={705--718},
  year={2025},
  publisher={IEEE}
}

@inproceedings{libritts,
  title     = {LibriTTS: A Corpus Derived from LibriSpeech for Text-to-Speech},
  author    = {Heiga Zen and Viet Dang and Rob Clark and Yu Zhang and Ron J. Weiss and Ye Jia and Zhifeng Chen and Yonghui Wu},
  year      = {2019},
  booktitle = {Interspeech 2019},
  pages     = {1526--1530},
  doi       = {10.21437/Interspeech.2019-2441},
  issn      = {2958-1796},
}

@inproceedings{utmos,
  title     = {UTMOS: UTokyo-SaruLab System for VoiceMOS Challenge 2022},
  author    = {Takaaki Saeki and Detai Xin and Wataru Nakata and Tomoki Koriyama and Shinnosuke Takamichi and Hiroshi Saruwatari},
  year      = {2022},
  booktitle = {Interspeech 2022},
  pages     = {4521--4525},
  doi       = {10.21437/Interspeech.2022-439},
  issn      = {2958-1796},
}

@inproceedings{ezvc,
  title={EZ-VC: Easy Zero-shot Any-to-Any Voice Conversion},
  author={Joglekar, Advait and Singh, Divyanshu and Bhatia, Rooshil Rohit and Umesh, Srinivasan},
  booktitle={Findings of the Association for Computational Linguistics: EMNLP 2025},
  pages={19768--19774},
  year={2025}
}

@inproceedings{sap,
  title     = {{The Interspeech 2025 Speech Accessibility Project Challenge}},
  author    = {Xiuwen Zheng and Bornali Phukon and Jonghwan Na and Ed Cutrell and Kyu J. Han and Mark Hasegawa-Johnson and Pan-Pan Jiang and Aadhrik Kuila and Colin Lea and Bob MacDonald and Gautam Mantena and Venkatesh Ravichandran and Leda Sari and Katrin Tomanek and Chang D. Yoo and Chris Zwilling},
  year      = {2025},
  booktitle = {{Interspeech 2025}},
  pages     = {3269--3273},
  doi       = {10.21437/Interspeech.2025-566},
  issn      = {2958-1796},
}

@inproceedings{l2arctic,
  title     = {{L2-ARCTIC: A Non-native English Speech Corpus}},
  author    = {Guanlong Zhao and Sinem Sonsaat and Alif Silpachai and Ivana Lucic and Evgeny Chukharev-Hudilainen and John Levis and Ricardo Gutierrez-Osuna},
  year      = {2018},
  booktitle = {{Interspeech 2018}},
  pages     = {2783--2787},
  doi       = {10.21437/Interspeech.2018-1110},
  issn      = {2958-1796},
}

@inproceedings{ecapa,
  title={{ECAPA-TDNN: Emphasized Channel Attention, propagation and aggregation in TDNN based speaker verification}},
  author={Desplanques, Brecht and Thienpondt, Jenthe and Demuynck, Kris},
  booktitle={Interspeech 2020},
  pages={3830--3834},
  year={2020}
}

@INPROCEEDINGS{unitdsr,
  author={Wang, Yuejiao and Wu, Xixin and Wang, Disong and Meng, Lingwei and Meng, Helen},
  booktitle={ICASSP 2024 - 2024 IEEE International Conference on Acoustics, Speech and Signal Processing (ICASSP)}, 
  title={UNIT-DSR: Dysarthric Speech Reconstruction System Using Speech Unit Normalization}, 
  year={2024},
  volume={},
  number={},
  pages={12306-12310},
  doi={10.1109/ICASSP48485.2024.10446921}}

@inproceedings{diffdsr,
  title     = {{DiffDSR: Dysarthric Speech Reconstruction Using Latent Diffusion Model}},
  author    = {Xueyuan Chen and Dongchao Yang and Wenxuan Wu and Minglin Wu and Jing Xu and Xixin Wu and Zhiyong Wu and Helen Meng},
  year      = {2025},
  booktitle = {{Interspeech 2025}},
  pages     = {2113--2117},
  doi       = {10.21437/Interspeech.2025-1770},
  issn      = {2958-1796},
}

@article{accent,
  title={SpeechAccentLLM: A Unified Framework for Foreign Accent Conversion and Text to Speech},
  author={Zhuangfei Cheng and Guangyan Zhang and Zehai Tu and Yangyang Song and Shuiyang Mao and Xiaoqi Jiao and Jingyu Li and Yiwen Guo and Jiasong Wu},
  journal={ArXiv},
  year={2025},
  volume={abs/2507.01348},
  url={https://api.semanticscholar.org/CorpusID:280149410}
}

@article{ln,
  title={Layer normalization},
  author={Ba, Jimmy Lei and Kiros, Jamie Ryan and Hinton, Geoffrey E},
  journal={arXiv preprint arXiv:1607.06450},
  year={2016}
}

@article{gelu,
  title={Gaussian error linear units (gelus)},
  author={Hendrycks, Dan and Gimpel, Kevin},
  journal={arXiv preprint arXiv:1606.08415},
  year={2016}
}

\end{document}